\documentclass[12pt]{article}
\usepackage{epsf}
\usepackage{amsfonts} %to get blackboard bold, bold etc
\usepackage{amsbsy} %likewise

\newdimen \halftextwidth
\newdimen\hsgraph \newdimen\vsgraph
\hsgraph=\hsize \advance\hsgraph by-1.2truein
\vsgraph=\vsize \advance\vsgraph by-1.5truein

\newcommand{\C}[1]{{\mathcal #1}}
\newcommand{\beq}[1]{\begin{equation}\label{#1}}
\newcommand{\eeq}{\end{equation}}
\newcommand{\bea}{\begin{eqnarray}}
\newcommand{\eea}{\end{eqnarray}}
\newcommand{\rf}[1]{(\ref{#1})} %sets the style for refering to equations

\newcommand{\half}{{1\over 2}}

\newcommand{\nn}{\nonumber}

\begin{document}
\topmargin 0pt
\oddsidemargin 5mm
\headheight 0pt
\topskip 0mm

\addtolength{\baselineskip}{0.20\baselineskip}

\pagestyle{empty}

\begin{flushright}
OUTP-97-14P\\
hep-th/9703141\\
20th March 1997
\end{flushright}

\begin{center}

\vspace{18pt}
{\Large \bf ND Tadpoles as New String States and Quantum Mechanical
 Particle-Wave Duality from  World-Sheet T-Duality}

\vspace{2 truecm}

{\sc Ian I. Kogan\footnote{e-mail: i.kogan1@physics.ox.ac.uk} 
and John F. Wheater\footnote{e-mail: j.wheater1@physics.ox.ac.uk}}

\vspace{1 truecm}

{\em Department of Physics, University of Oxford \\
Theoretical Physics,\\
1 Keble Road,\\
 Oxford OX1 3NP, UK\\}

\vspace{3 truecm}

\end{center}

\noindent
{\bf Abstract.} We consider  new  objects in bosonic open string theory --
 ND tadpoles, which have  N(euman) boundary conditions at one end of 
 the world-sheet and D(irichlet) at the other, 
must exist due to  $s-t$ duality
 in a string theory with both NN strings and D-branes.
  We demonstrate how to interpolate
 between N and D boundary conditions. In the case of mixed boundary
 conditions the  action for a quantum particle is induced on the
boundary. Quantum-mechanical
 particle-wave duality, a dual description of a quantum
particle in either the  coordinate or the  momentum representation,
is induced by  world-sheet T-duality. The famous
 $\tilde{R} = \alpha/R$ relation is equivalent to the quantization
 of the  phase space area of a  Planck cell $\oint pdx = 2\pi\hbar$. 
 We also introduce a boundary operator $\C Z$ - a ``Zipper'' which
changes the boundary condition from N into D and vice versa.

\vfill
\begin{flushleft}
\end{flushleft}
\newpage
\setcounter{page}{1}
\pagestyle{plain}

%A few examples of fonts in maths.......
%\beq \B R\quad \B Z \quad\boldsymbol\alpha \quad\C Z\label{first}\eeq

%And then a postscript picture....
%\begin{figure}[h]
%{\epsfxsize=\textwidth \epsfbox{diags.eps}}
%\caption{The vacuum diagrams corresponding to the 2nd, 4th and 6th
%connected vacuum correlation functions respectively.}
%\end{figure}

Recently the theory of open strings, which was overshadowed for 
many years by the explosion of interest in
closed strings, has attracted renewed interest largely because of
the discovery of D-branes \cite{Dbranes, Tassi}.  Let $\sigma\in[0,\pi]$ 
be the normal
coordinate  and $\tau$ be the tangential
coordinate on an open string world-sheet (fig.1) and define
$z=e^{\tau+i\sigma}$.  A normal open 
string configuration $X^\mu(z,\bar z)$ has Neuman boundary conditions,
$\partial_\sigma X^\mu=0$ at $\sigma=0,\pi$. On the other hand a
D-brane has Dirichlet boundary conditions $\partial_\tau X^\mu=0$ 
at $\sigma=0,\pi$.  However, because an open string has two boundaries,
one at $\sigma=0$ and one at $\sigma=\pi$ it is also possible to have 
Dirichlet boundary conditions at one boundary and Neuman at the other.
The respective mode expansions are given by (for a review see \cite{Tassi}
and references therein)
\bea
NN&X^\mu(z,\bar z)=&x^\mu-i'\alpha'p^\mu\ln z\bar z +i\sqrt{\alpha'\over2}
\sum_{m\ne0}{\alpha_m^\mu\over m}\left(z^{-m}+\bar z^{-m}\right)\nn\\
DD&X^\mu(z,\bar z)=&-i{\delta X^\mu\over 2\pi}\ln {z\over\bar z}
 +i\sqrt{\alpha'\over2}
\sum_{m\ne0}{\alpha_m^\mu\over m}\left(z^{-m}-\bar z^{-m}\right)\nn\\
ND,DN&X^\mu(z,\bar z)=&i\sqrt{\alpha'\over2}
\sum_{m}{\alpha_m^\mu\over m}\left(z^{-m-\half}\pm\bar
z^{-m-\half}\right)
\label{1}
\eea
 although, as we will argue later, the expression for ND is incomplete.
\begin{figure}[b]
{\epsfxsize=\halftextwidth \epsfbox{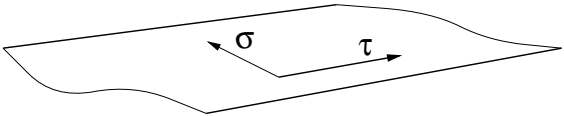}}
\caption{The world sheet coordinates $\sigma$ and $\tau$.}
\end{figure}
NN corresponds to
 open strings with free ends (which move at the speed of light).
DD strings are open strings attached 
to a fixed surface or point -- the D-brane (in this case
$\delta X^\mu$ is the displacement
between two endpoints of an open string).
ND strings, we will call them tadpoles, correspond to an open string which
is attached at one end to a D-brane while the other end is free.

ND boundary conditions have been studied earlier, both in string
theory \cite{siegel,sagnot,mbg} and
in  $c=1$ boundary conformal field theory \cite{CKPT}. 
 Here we take the new  step
of stressing that
the ND tadpoles must be considered on an equal physical footing with 
the usual NN open strings and the D-branes.  By analogy with what occurs
in ordinary field theory it is common to think of the NN strings as the 
fundamental quanta, $f$, and of the D-branes as the solitons, $S$, of the
theory; however we then expect the process
\beq{2}
f\bar f \to S\bar S
\eeq
to occur. There are two possible ways to describe this process in
string theory.  
The first one is through the closed string world-sheet 
 interpolating between N and D closed boundaries (Fig.2a). 
  One can look at this either   as a transition between N and D boundaries 
via the closed string or as a loop of an ND tadpole (Fig.2b).
 In this case the quantum numbers of the initial $f\bar f$ pair
 (as well as the final $ S\bar S$ pair) must be compatible with the quantum
numbers of closed string states (graviton, dilaton, etc). 
\begin{figure}[t]
{\epsfxsize=\textwidth \epsfbox{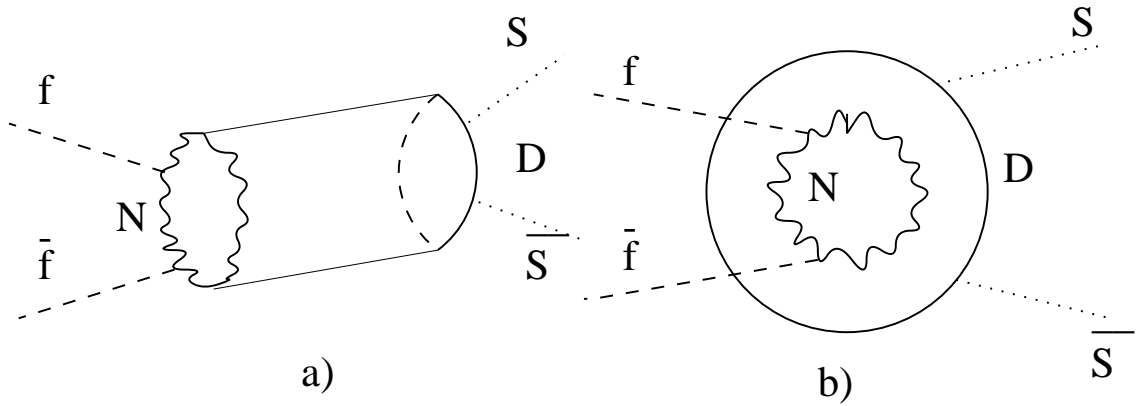}}
\caption{a) Soliton pair production from fundamental quanta through a
closed string intermediate state and b) the 
corresponding t-channel process contains an ND tadpole in the intermediate
state.}
\end{figure}
 When this is not 
true, for example if instead of $e^+ e^-$
annihilation into monopole-antimonopole  pair one has $\bar{\nu}e^-$
 annihilation into  monopole-antidyon pair, 
 it is impossible to  connect initial and final states by  a closed
 string  state.  In this case
the world sheet for this process must look like Fig.3a.
We have introduced an operator $\C Z$ which we will call the ``Zipper''
which acts on the boundary to change N boundary conditions into D (there
is correspondingly an anti-zipper which changes D to N).  In the $t$-channel
the diagram becomes Fig. 3b 
which is nothing but the propagation of an ND tadpole in the 
intermediate state. We conclude that in a theory able to describe the
transition between fundamental quanta and solitons the ND tadpoles must
be present because of standard $s-t$ crossing symmetry.  ND states also
appear as a consequence of the modular invariance of NN and DD systems
as was first observed in \cite{sagnot} where the one-loop partition
function was studied on open string orbifolds.
\begin{figure}[t]
{\epsfxsize=\halftextwidth \epsfbox{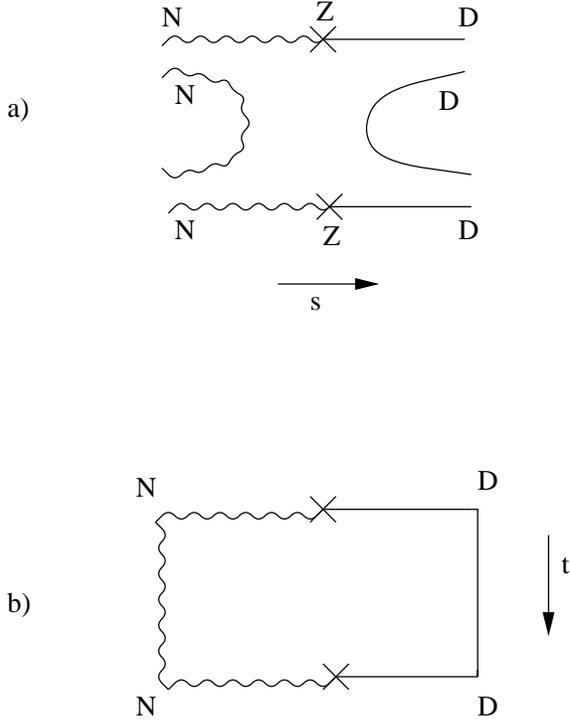}}
\caption{Production of $S\bar S$ through an open string intermediate
state in a) the s-channel and b) the t=channel. Crosses mark the Zipper.}
\end{figure}
Note  that even in the first case of Fig.2a the Zipper
is a natural object. The problem is that if it is necessary to produce
two $D$-branes at two different points in space (for example in 
pair production other than at threshold) it is very hard to
 imagine a boundary condition along the D boundary where $X$ takes
one value $X_1$ on one part of the boundary and then suddenly is 
changed into $X_2$ along  another part. This can be done easily by
 inserting a Zipper - un-Zipper pair  $\C Z- \bar{\C Z}$
 which will transform D conditions
with boundary value $X_1$ into N conditions and then back into D
conditions but with a new value $X_2$.

Strings originated in the study of hadronic physics; more 
recently models of hadrons  motivated by QCD in which quarks are bound
together by tubes of non-abelian flux have been developed  (see
 \cite{ovw} and references therein).
  Using the
language of quarks and hadrons one can say that NN strings are  $q\bar q$
meson 
states with massless quarks, DD are $Q\bar Q$ quarkonia systems with 
infinitely heavy quarks and ND tadpoles are the $Q\bar q$ mesons --
the ``B'' mesons, or heavy-light mesons, of QCD. 
 Interestingly even
baryons  in QCD can be described in the 
same spirit 
as a combination of three ND tadpoles 
with a common D boundary which is the centre of mass;  once again the baryon
is a sort of soliton ( see fig.4b which actually
shows a SU(4) baryon for technical reasons). The center of mass, sometimes called the string
junction, is a soliton collective coordinate. The dynamics of this
object  was discussed in \cite{junction}.
The Zipper operator is the key to describing
weak decays of mesons; its ability to turn D into N parallels the quark
process  $Q\bar Q\to Q\bar q +W$
(fig.4a);
 in the stringy language the $W$ vertex must contain  $\C Z$.

\begin{figure}[h]
{\epsfxsize=\textwidth \epsfbox{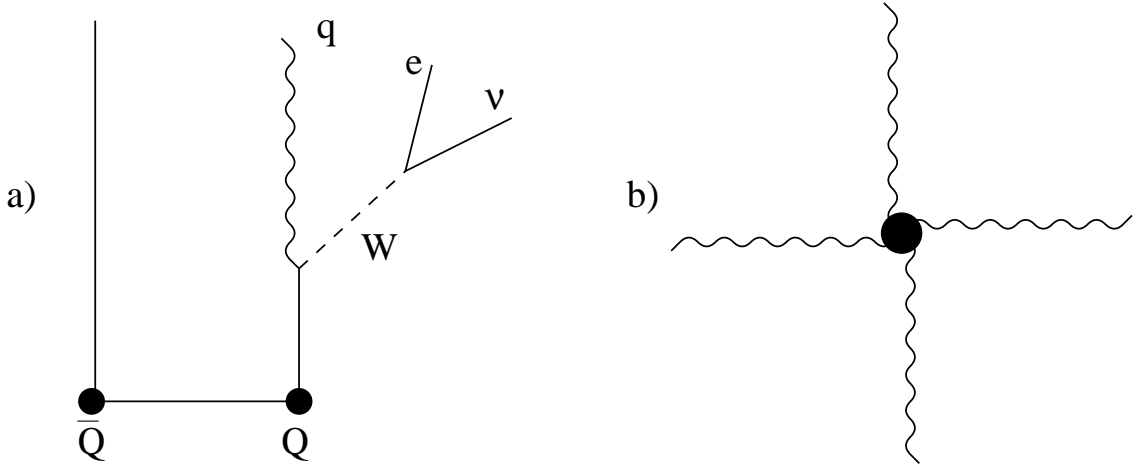}}
\caption{a) Semi-leptonic weak decay of heavy quarkonia and 
 b) An $N_c=4$ baryon.}
\end{figure}

The existence of $\C Z$ which changes N to D raises the question of the
interpolation between N and D. A more general boundary condition is
\beq{3}
f(\tau)\partial_\sigma X^\mu+X^\mu=const
\eeq
If $f=\infty$ we get N and if $f=0$ we get D boundary conditions; however
the more general condition \rf{3} is not consistent with the unmodified
string action which is, in the conformal gauge,
\beq{4}
{\C S}=\half\int_\Sigma (\partial_aX^\mu\partial_aX^\mu)\,d^2\xi
\eeq
The variation of $\C S$ under $X\to X+\delta X$ is
\beq{5}
\delta{\C S}=-\int_\Sigma \delta X^\mu\partial^2X^\mu\,d^2\xi
+\int_{\partial\Sigma}\delta X^\mu\partial_\sigma X^\mu\,d\tau
\eeq
and the condition for an extremum  is that either $\delta X^\mu=0$ or
$\partial_\sigma X^\mu=0$ on the boundary. To get \rf{3} we must add an
extra term to the action and consider
\beq{6}
{\C S}'=\half\int_\Sigma (\partial_aX^\mu\partial_aX^\mu)\,d^2\xi+
\half\int_{\partial\Sigma}f(\tau)(\partial_\sigma X^\mu)^2\,d\tau
\eeq
whose variation is
\beq{7}
\delta{\C S}=-\int_\Sigma \delta X^\mu\partial^2X^\mu\,d^2\xi
+\int_{\partial\Sigma}(\delta X^\mu+f(\tau)\,\delta(\partial_\sigma X^\mu))
 \,\partial_\sigma X^\mu\,d\tau
\eeq
Now we have an extremum when $X^\mu$ satisfies Laplace's equation and
the boundary condition
\beq{8}
\delta( X^\mu+f(\tau)\,\partial_\sigma X^\mu)=0
\eeq
which is the same as \rf{3}.

To understand the meaning of this extra term in the action we shall make
a T-duality transformation; this maps $X$ to a new field $\Phi$ with the
relation
\beq{9}
\partial_i X^\mu =\epsilon_{ij}\partial_j\Phi^\mu
\eeq
With this change of variables ${\C S}'$ becomes
\beq{10}
\tilde{\C S}'=\half\int_{\tilde \Sigma}(\partial_a\Phi^\mu)^2\,d^2\xi+
\half\int_{\partial\tilde\Sigma} f(\tau)
(\partial_\tau\Phi^\mu)^2\,d\tau
\eeq
Now the boundary term resembles the action for the path integral describing
the motion of a free particle along the trajectory $\Phi^\mu(\tau)$.
 By considering the variation 
$\Phi\to \Phi+\delta \Phi$ we find 
that $\Phi$ satisfies Laplace's equation but
this time with the boundary conditions
\beq{11}
\partial_\sigma\Phi^\mu -\partial_\tau(f(\tau)\partial_\tau\Phi^\mu)=0
\eeq
Using \rf{9} we can write this in terms of $X$ as
\beq{12}
\partial_\tau( X^\mu+f(\tau)\,\partial_\sigma X^\mu)=0
\eeq
which is the same as \rf{3}.
If $f=\infty$ we get D and if $f=0$ we get N boundary conditions for 
$\Phi^\mu$.

 We see that  by introducing  mixed boundary conditions we have
induced some dynamics on the world-line formed by the boundary of the
world-sheet.  This  is the same
phenomenon as in the  case of a three-dimensional topologically massive
gauge theory which induces (depending on boundary conditions) a
non-trivial dynamics on a three-dimensional boundary (see for example
\cite{ck} and references therein).  It is very tempting to describe
 this induced boundary dynamics as a massive relativistic particle
with mass proportional to $f$. 
  Let us demonstrate that this is precisely the case.

 Consider the   Feynman path
integral representation for the relativistic  propagator 
of a scalar particle of mass $M$  in $d$ space-time dimensions 
(for simplicity consider a Wick rotated
 propagator in the Euclidean
region)
\begin{eqnarray}
G(x,y) = \int \frac{dp}{(2\pi)^{d}} \frac{e^{ip(x-y)}}
{p^{2} + M^{2}} =
\frac{1}{2}\int \frac{dp}{(2\pi)^{d}} e^{ip(x-y)}\int_{0}^{\infty} 
dT e^{-(p^{2}+M^{2})T/2}=
\nonumber \\ \\
= \frac{1}{2}\int_{0}^{\infty} dT e^{-M^{2}T/2} 
{\displaystyle{\int_{x(0) = x}^{x(T) = y}}} 
{\cal D}x(\tau)
\exp[-\frac{1}{2}\int_{0}^{T}d\tau \dot{x}^{2}] 
\;\;\;\;\;\;\;\;\;\;\;\;\nonumber
\end{eqnarray}
It is possible to represent the integration over the
 proper time $T$ as 
an integration over one-dimensional metrics $h(\tau)$ modulo
 the one-dimensional
reparametrization group $f(\tau)$ (see, for example,
 \cite{book}), so the
relativistic scalar propagator takes the form
\begin{eqnarray}
G(x,y) =  \int
{\cal D}x(\tau){\cal D}h(\tau)
\exp\left[-\frac{1}{2}\int d\tau  \frac{\dot{x}^{\mu}\dot{x}_{\mu}}
{\sqrt{h(\tau)}} - \frac{1}{2}M^{2}
\int d\tau \sqrt{h(\tau)}\right]
 =   ~~~ \nonumber \\ \label{pathintegral}\\
 \int
{\cal D}x(\tau){\cal D}p(\tau) {\cal D}h(\tau) 
\exp\left[i\int  d\tau p^{\mu}(\tau)\dot{x}_{\mu}(\tau)
 -\frac{1}{2}\int d\tau
\sqrt{h(\tau)}
 \left( p^{\mu}(\tau)p_{\mu}(\tau) + M^{2} \right)
\right] ~~
\nonumber
\end{eqnarray}
 where the coordinate along the world-line is  
$\tau$ and $h(\tau)$ is the
 one-dimensional
metric. The proper time is the only  reparametrization 
invariant
 characteristic of the metric -- the length of the path
 $\int_{0}^{1}\sqrt{h}d\tau = T$.  Let us now change the variable
 $\tau$ into a new time $t(\tau)$ in such a way that
\begin{equation}
\frac{dt}{d\tau} = M^2\sqrt{h(\tau)}
\label{15}
\end{equation}
after which the action along the trajectory $x^\mu (\tau)$  in the first
 equation of (\ref{pathintegral}) can be rewritten as
\begin{equation}
\frac{1}{2}M^2\int dt   \frac{dx^{\mu}}
{dt}\frac{dx_{\mu}}
{dt} - \frac{1}{2}\int dt
\label{16}
\end{equation}
 where the second term is some constant and can be dropped. 
Comparing this with (\ref{10}) one can see that
$\Phi^\mu$ describes the  target-space coordinate 
of a massive particle and that
 $f$  is indeed
 related to $M$ (actually $M^2$) by
\begin{equation}
f(\tau) \frac{dt}{d\tau} = M^2
\label{17}
\end{equation}
But if $\Phi^\mu$ describes the  target-space coordinate 
of a massive particle, what is the role of its dual field $X^\mu$ ?

 To answer this question we have to rewrite the boundary action (\ref{6})
 not in terms of $\partial_\sigma X$ but in terms of $X$. Fortunately
we can do this using the boundary condition (\ref{3})
 ~   $X = - f\partial_\sigma
 X$ ~ where $X$ in (\ref{3}) is shifted  to put  $const = 0$.
  Then the action can be written as
\begin{equation}
\half\int_{\partial\Sigma}\frac{1}{f(\tau)} X^\mu(\tau)
 X_\mu(\tau)   \,d\tau
= \frac{1}{2 M^2}
\int_{\partial\Sigma} X^\mu(t) X_\mu(\tau) \,dt
\end{equation}
 where  we used (\ref{17}) again.  Comparing this with the second
 line in (\ref{pathintegral})  and using (\ref{15}) we
immediately  see that  $X^{\mu} (\tau)$  is the  momentum $p(\tau)$.

 This is a very remarkable fact - we just demonstrated that the 
   coordinates $X^{\mu}$ and $\Phi^{\mu}$ are the coordinates in 
 {\it phase space}   and that the T-duality transformation on the
 world-sheet induces a transformation in  a  target- phase space!
  In other words T-duality is related to de Broglie  quantum-mechanical
 ``particle - wave''  duality.

  Note that by definition T-duality relates ``coordinates''
 for $X$  and ``momenta''  for $\Phi$ and vice versa because of the
 relation  (\ref{9}) -- indeed the spatial derivative (``coordinate'')
  of one of the fields is related to  the 
 time derivative (``conjugate momentum'')    of  the other one. However
it is important to stress that what we have just found is much more
amusing  -- we obtained a  quantum description of physics 
in  target space, even in the
limit when the string coupling is zero. Naively we must have 
classical physics in  target space; nevertheless the world-sheet
  already knows about quantum reality.

The relationship between  compactification radii for $X$ and $\Phi$,
$R \rightarrow
\tilde{R}= 
 \alpha'/R$, implies  the famous quantization
condition  of  a Planck cell in  phase space:
 \begin{equation}
 \oint p dx = 2\pi \hbar
\end{equation}
This arises because
  $X$ and $\Phi$ (or,  more precisely, 
their zero modes)  play the roles of
  phase space  coordinates $p$ and $x$;
 recalling that there
 is a $1/2\pi \alpha'$ factor in front of the world-sheet action
we have
\begin{equation}
x = \Phi; ~~~~ p = \frac{\hbar}{2\pi\alpha'} X
\end{equation}
 Taking into account that  $\Phi \in (0, 2\pi R)$ and
$X \in (0, 2\pi\tilde{R})$ we can easily see that
\begin{eqnarray}
\oint pdx =\frac{\hbar}{2\pi\alpha'} (2\pi R) ( 2\pi\tilde{R}) =
(2\pi\hbar) \frac{R\tilde{R}}{\alpha'} \nonumber \\
\oint pdx = 
(2\pi\hbar) \Rightarrow \tilde{R} = \frac{\alpha'}{R}
\label{tduality}
\end{eqnarray}

 Now we can write down an expression for the Zipper operator --
it is nothing but
 the string generalization of the $p \dot{x}$ term in the path integral
 (\ref{pathintegral}).  One can define it as (we introduce explicitly the
factor $1/2\pi\alpha'$)
\begin{eqnarray}
\C Z = \int {\C D}X(\tau){\cal D}\Phi(\tau)
\exp\left[\frac{i}{2\pi\alpha'}\int \, d^2 \xi \,\epsilon_{ab}\partial_a X
 \partial_b \Phi\; 
\right]
\label{zipper}
\end{eqnarray}
 and this is precisely what changes N into D - but as we have already
seen this is nothing but a change of the quantum-mechanical
description
 from the $x$- into the  $p$- representation. The density $\epsilon_{ab}\partial_a X
 \partial_b \Phi$ is a total derivative so the exponential factor is
reduced to an integral along the boundary 
\begin{equation}
\frac{1}{2\pi\alpha'} \int  d\tau X(\tau)\dot{\Phi}(\tau)
 \end{equation}
which corresponds to $\int  d\tau  p \dot{x}$ term.  To change
boundary conditions from N to D at some point $z$ on the boundary
 we have to provide the boundary conditions
 for fields $X$ and $\Phi$ such that $X = 0$ for all $\tau < z$
 and $\Phi = 0$ for all $\tau > z$; for the un-Zipper $\bar{\C
Z}$ one has to exchange  boundary conditions for $X$ and $\Phi$.
 Let us  note that due to the condition (\ref{tduality}) the
world-sheet action  in the definition of the Zipper operator
 does not depend on the homotopy class of the maps $X(\xi)$ and $\Phi(\xi)$
-- if one has a map where both $X$ and $\Phi$ wrap around the
respective circles with circumferences  $2\pi R$ and $2 \pi
\tilde{R}$, there will be an extra  factor 
\begin{equation}
\exp\left[2\pi\; i \;\frac{\tilde{R} R}{\alpha'} \right]
 = 1
\end{equation}
 which means that the operator (\ref{zipper}) is well-defined.

  One can use these operators to define  vertex operators at
world-sheet boundaries with   D  boundary conditions. Usually it is
 impossible to transfer any momentum at the D boundary  because the
 standard exponential operator $\exp(i k_{\mu} X^{\mu})$ does not
exist as a fluctuating field along the boundary.
 The only way to transfer  momentum to the D-brane was through
the bulk vertex operators, i.e. we could scatter  closed strings,
but not open ones, off the D-brane. Now we can  do something new, namely insert
 a pair $\C Z -  \bar{\C Z}$ at the D boundary and create a short
 interval  with N boundary conditions, where we can insert any vertex
 operator $V_{N}(\tau)$. Shrinking the size of the N interval to zero
 we can define  new vertex operators  which 
can then be integrated  along the D boundary 
\begin{equation}
 \int d\tau V_{D}(\tau) = \lim_{\epsilon \rightarrow 0}
\int d\tau \bar{\C Z}(\tau + \epsilon) V_{N}(\tau)
\C Z (\tau - \epsilon)
\end{equation}
 These new vertex operators will  enable us
 to study the  dynamics of D-branes.

A further indication of non-trivial connections between target space
and world-sheet induced by T-duality comes from considering the ND tadpoles.
 In  the case of NN and DD strings  one can either  study
T-duality,  in which case NN and DD strings are interchanged,
 or a world-sheet parity transformation $\sigma\to \pi-\sigma$
 which  exchanges the boundaries and leaves both  types of strings
intact.  It seems that T-duality and  world-sheet parity
transformation are unrelated, but this is wrong.
   It is quite clear that  both
T-duality and a world-sheet parity transformation
 transform the   ND  tadpole  into  the DN tadpole and vice versa.  
Let 
\bea
X_+^\mu(z)&=&x_+^\mu+i\sqrt{\alpha'\over2}
\sum_{m}{\alpha_m^\mu\over m}\left(z^{-m-\half}\right)\nn\\
X_-^\mu(\bar z)&=&x_-^\mu+i\sqrt{\alpha'\over2}
\sum_{m}{\alpha_m^\mu\over m}\left(\bar z^{-m-\half}\right)
\eea
Then a tadpole with N boundary conditions at $\sigma=0$ and
D boundary conditions at $\sigma=\pi$ is
\bea
X^\mu(z,\bar z)&=&X_+^\mu(z)+X_-^\mu(\bar z)\nn\\
&=&x_+^\mu+x_-^\mu+i\sqrt{\alpha'\over2}
\sum_{m}{\alpha_m^\mu\over m}\left(z^{-m-\half}+\bar z^{-m-\half}\right)
\eea
A world-sheet parity transformation $\sigma\to \pi-\sigma$ transforms this
to a tadpole with D boundary conditions at $\sigma=0$ and
N boundary conditions at $\sigma=\pi$
\bea
{\C P}X^\mu(z,\bar z)
&=&x_+^\mu+x_-^\mu+i\sqrt{\alpha'\over2}
\sum_{m}{\alpha_m^\mu\over m}(-1)^m i\left(\bar z^{-m-\half}- z^{-m-\half}\right)
\eea
However a T transformation also has the same effect giving
\bea
\Phi^\mu(z,\bar z)&=&X_+^\mu(\bar z)-X_-^\mu( z)\nn\\
&=&x_+^\mu-x_-^\mu+i\sqrt{\alpha'\over2}
\sum_{m}{\tilde{\alpha}_m^\mu\over m}
\left(\bar z^{-m-\half}- z^{-m-\half}\right)
\eea
where the dual oscillators $\tilde{\alpha}_m = i(-1)^m {\alpha}_m$.
Note  that this ND tadpole is in a different position in target space
($x_+^\mu-x_-^\mu$ rather than $x_+^\mu+x_-^\mu$); the commutator of
a T transformation and a world sheet parity $\C P$ operation 
is a target space translation.  Thus we have to add to the ND mode expansion
 in (\ref{1}) a zero mode term which reflects the fact that the
center of mass of the ND tadpole can be anywhere in target space.

From another point of view 
the mixed boundary conditions \rf{3} generate a world
sheet boundary state with mass $f^{-1}$; 
this is easily seen by considering the
mode expansion which yields a solution
\beq{last}
X^\mu(z,\bar z)=x_+^\mu+x_-^\mu+ip^\mu\log{z\over\bar z}+i\sqrt{\alpha'\over2}
\sum_{m}{\alpha_m^\mu\over m}\left(z^{-m}+{1-ifm\over 1+ifm}\bar z^{-m}\right)
\eeq
As usual the dual field is obtained by considering,
instead of the sum of left and right movers 
$ X_+^\mu(z) + X_-^\mu(\bar z)$, the difference 
 $X_+^\mu(\bar z)-X_-^\mu(z)$ with the
 simultaneous exchange of $z$ and $\bar{z}$. 
 The appearance of a pole at $m=if^{-1}$ in \rf{last} means that the world
sheet two point function $\langle X X \rangle$ 
gains a contribution decaying exponentially with (world sheet) 
distance from the boundary; there is a massive state
living on the boundary. It is interesting to note that this mass is, up to a
factor of $\alpha'$, the inverse of the target space mass we found above;
this suggests that the world sheet effect should be regarded as resulting
from the gravitation induced by the target space mass, i.e. 
comparing world-sheet and target-space scales (in light-cone
gauge, for example), then this scale is not a Compton  length
 $\lambda_{C} = \hbar / M$, but a  gravitational (Schwarzschild)
 radius $\lambda_{G} \sim \alpha' M$. The dynamics of this boundary
state is very interesting and we hope to discuss it further in future
publications. Recently the boundary states in one-dimensional  Hubbard
 and t-J models were considered in \cite{essler} where their existence
 was related to the mixed boundary conditions.

 There are  several interesting  problems we hope to address in the
future.  The first one is to include 
 world-sheet gravity. In this case the one-dimensional boundary action
 will be equipped with a naturally induced metric $h(\tau)$ and the integral
 over this metric will arise as a part of a path integral over the 2d
metric $g_{ab}(\xi)$ on a world-sheet with a boundary. It is an
interesting  problem to study the interaction of the boundary state with the
 2d  and induced 1d gravity and,  for example, to see how the
mixed boundary conditions  will affect the conformal anomaly. It is quite
 possible that there will be boundary states in the gravitational sector
too.   Another
problem is to include supersymmetry and world-sheet fermions and to 
demonstrate that the action for a supersymmetric quantum-mechanical
 particle will be induced at the boundary. It is also 
 interesting  to look at  mixed boundary conditions and
 induced boundary dynamics in rigid strings \cite{book}.

\vspace{1 truecm}
\noindent We acknowledge valuable conversations with J. Cardy, F. Essler,
 J. Paton, A. Sen  and especially  N. Mavromatos.

\end{document}